\documentclass{article} 
\usepackage{iclr2025_conference,times}

\usepackage{amsmath,amsfonts,bm}

\def\eqref#1{equation~\ref{#1}}

\def\1{\bm{1}}

\DeclareMathAlphabet{\mathsfit}{\encodingdefault}{\sfdefault}{m}{sl}
\SetMathAlphabet{\mathsfit}{bold}{\encodingdefault}{\sfdefault}{bx}{n}

\usepackage{hyperref}
\usepackage{url}
\usepackage{times}
\usepackage{latexsym}
\usepackage{booktabs}
\usepackage{multirow}
\usepackage{etex}
\usepackage{amsmath}
\usepackage{diagbox}
\usepackage{enumitem}
\usepackage{amssymb}
\usepackage{scalerel}
\usepackage{subcaption}
\usepackage{microtype}
\usepackage{inconsolata}
\usepackage{graphicx}
\usepackage{xcolor}
\usepackage{colortbl}
\usepackage{xspace}
\usepackage{tikz}
\usepackage[normalem]{ulem}

\newcommand{\todoc}[2]{{\textcolor{#1}{\textbf{#2}}}}

\renewcommand{\todoc}[2]{\relax}

\newcommand{\rev}[1]{{#1}}

\newcommand{\code}[1]{\texttt{\small #1}} 
\newcommand{\ours}{Nova}
\newcommand{\nocl}{Nova$_{\footnotesize -CL-HA}$}
\newcommand{\noha}{Nova$_{\footnotesize -HA}$}
\newcommand{\nofcl}{Nova$_{\footnotesize -FCL-HA}$}
\newcommand{\noocl}{Nova$_{\footnotesize -OCL-HA}$}

\definecolor{lightblue}{HTML}{CCE5FF}
\definecolor{lightgreen}{HTML}{ACD941} 
\definecolor{darkgreen}{HTML}{38B178} 
\definecolor{yellow}{HTML}{FDE725} 
\definecolor{lightgrey}{rgb}{0.9,0.9,0.9}

\newcommand{\distance}{10pt}
\title{Nova: Generative Language Models for Assembly Code with Hierarchical Attention and Contrastive Learning}

\author{Nan Jiang \\
{\small Purdue University} \\
{\small jiang719@purdue.edu}
\And
Chengxiao Wang \\
{\small University of Illinois Urbana-Champaign} \\
{\small cw124@illinois.edu}
\And
Kevin Liu \\
{\small Lynbrook High School} \\
{\small kevin.bx.liu@gmail.com}
\And
Xiangzhe Xu \\
{\small Purdue University} \\
{\small xu1415@purdue.edu}
\And
Lin Tan \\
{\small Purdue University} \\
{\small lintan@purdue.edu}
\And
Xiangyu Zhang \\
{\small Purdue University} \\
{\small xyzhang@cs.purdue.edu}
\And
Petr Babkin \\
{\small J.P. Morgan AI Research} \\
{\small petr.babkin@jpmorgan.com}
}

\iclrfinalcopy 
\begin{document}
\maketitle

\begin{abstract}
Binary code analysis is the foundation of crucial tasks in the security domain; thus building effective binary analysis techniques is more important than ever. Large language models (LLMs) although have brought impressive improvement to source code tasks, do not directly generalize to assembly code due to the unique challenges of assembly: (1) the low information density of assembly and (2) the diverse optimizations in assembly code. To overcome these challenges, this work proposes a \emph{hierarchical attention} mechanism that builds attention summaries to capture the semantics more effectively, and designs \emph{contrastive learning objectives} to train LLMs to learn assembly optimization. Equipped with these techniques, this work develops \emph{\ours{}}, a generative LLM for assembly code. \ours{} outperforms existing techniques on binary code decompilation by up to 14.84 -- 21.58\% (absolute percentage point improvement) higher Pass@1 and Pass@10, and outperforms the latest binary code similarity detection techniques by up to 6.17\% Recall@1, showing promising abilities on both assembly generation and understanding tasks.
\end{abstract}

\section{Introduction}
Binary code plays an irreplaceable role in the security domain, being the foundation of crucial tasks including vulnerability detection~\citep{vul-motivate-1, vul-motivate-2, vul-motivate-3}, malware detection~\citep{malware-motivate-1, malware-motivate-2, malware-motivate-3}, binary recovery~\citep{codeart,osprey, dirty}, and legacy software maintenance~\citep{legacy-motivate-1, legacy-motivate-2, legacy-motivate-3}. 
For example, when performing tasks such as identifying attacks and malware, security analysts often only have access to assembly, i.e., the human-readable representation of binary code, which is extremely difficult to understand~\citep{codeart,osprey, dirty}. 
Thus, combined with the increasing sophistication of cybercrime that poses significant threats worldwide (e.g., cybercrime is predicted to cost the world \$10.5 trillion annually by 2025~\citep{cyber-attack}), effective binary analysis techniques are in high demand.

\begin{figure*}[htp]
    \centering
    \includegraphics[width=0.95\linewidth]{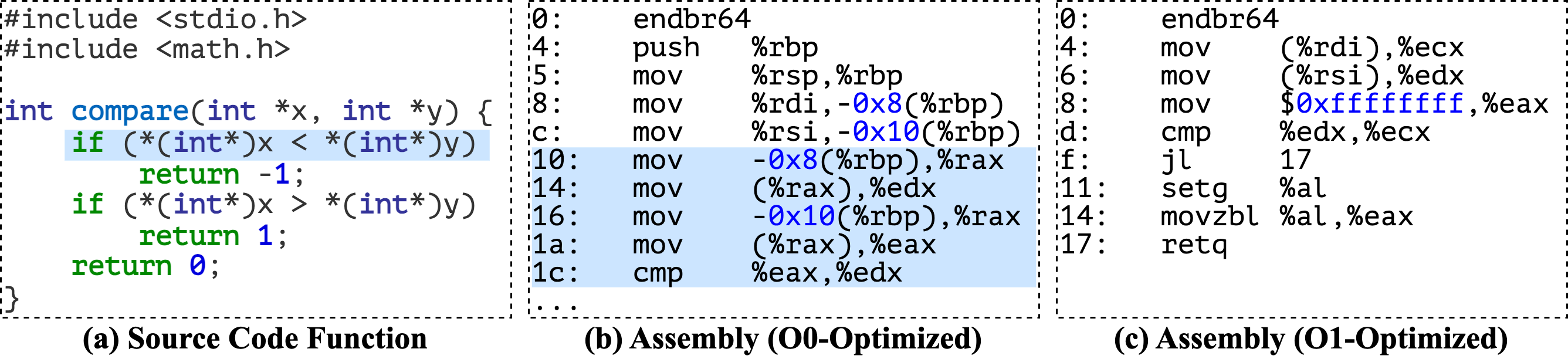}
    \caption{Example that shows the semantics and diverse optimizations of assembly code.}
    \label{fig:motivation}
\end{figure*}

Large language models pre-trained on source code have brought improvement in various software development domains~\citep{related-generation-1, related-generation-2, related-generation-3, related-generation-4, related-apr-1, related-apr-2}. However, these LLMs are not designed for or trained with assembly corpus, not achieving their full potential on binary code analysis tasks such as binary code similarity~\citep{jtrans, xu2023diemph}, malware detection~\citep{codeart}, and binary code decompilation~\citep{llm4decompile, slade, related-decompile-1}.

Existing work applying LLMs on assembly code mainly piggybacks on encoder-style LLMs~\citep{jtrans, codeart, xu2023diemph}, unable to benefit from the more extensive pre-training, updated architectures, scaling of state-of-the-art generative LLMs. Other work using generative LLMs for decompilation shows a low unit test passing rate of the decompiled programs~\citep{llm4decompile, slade}. 

The challenges of leveraging generative LLMs for assembly code are twofold. First,  compared to source code, assembly code has a \emph{lower information density}. 
A short source-code sequence maps to an assembly-code sequence that is often several times longer.  Thus, assembly semantics span across a \emph{long sequence of tokens}.   
Figure~\ref{fig:motivation} (a) shows an example of a source code function that compares two integers, while Figure~\ref{fig:motivation} (b) shows its corresponding assembly code optimized with \code{O0} flag. In the \code{O0}-optimized assembly code, the five instructions from \colorbox{lightblue}{\raisebox{0pt}[5pt][0pt]{\code{10:\;\;\;mov\;\;\;-0x8(\%rbp),\%rax}}} to \colorbox{lightblue}{\raisebox{0pt}[5pt][0pt]{\code{1c:\;\;\;cmp\;\;\;\%eax,\%edx}}} perform the checking whether the value of \code{x} is smaller than the value of \code{y} (correspond to \colorbox{lightblue}{\raisebox{0pt}[5pt][0pt]{\code{if (*(int*)x < *(int*)y)}}} in the source code). A single assembly instruction alone represents little meaningful semantics in the source code.  It is the combinations of \emph{many} \emph{instructions} and the \emph{dependencies} between them represent the semantics. Such combinations of instructions are long, which is hard for LLMs to learn. 

Second, assembly code is diverse due to compiler optimization. The assembly code of the same source code function looks dramatically different with different compiler optimization.
Figure~\ref{fig:motivation} (c) shows the assembly of the same function compiled with \code{O1} and \code{O0} flags, which consists of a significantly different set of instructions. Such syntax diversity is hard for LLMs to learn, preventing LLMs from obtaining consistently good performances on differently optimized assembly code.

In this work, we develop \ours{}, a generative foundation LLM pre-trained for assembly code with two key novelties. First, to address the low-information-density and long-sequence challenge, we design a hierarchical self-attention, which contains three categories of attention at different levels of granularity: intra-instruction attention, preceding-instruction attention, and inter-instruction attention. The key insight is to build \emph{attention summaries}, i.e., we create per-statement attention \emph{labels}, which act as the summary of a statement. We then use preceding-instruction attention to capture semantics between a token and its preceding instruction label and use inter-instruction attention for long dependencies. Besides, we design \emph{functionality contrastive learning} and \emph{optimization contrastive learning} objectives to train \ours{} to learn the semantics behind the diverse syntax of assembly.

This work makes the following contributions:
\begin{itemize}[leftmargin=10pt, noitemsep]
    \item We propose a novel hierarchical attention mechanism that captures the assembly's low-density semantics at three granularity levels.
    \item We design contrastive learning objectives to train LLMs to learn assembly with diverse optimizations and encode assembly more efficiently.
    \item We develop \ours{}, a generative foundation LLM with hierarchical attention and contrastive learning for assembly. \ours{} outperforms state-of-the-art on binary decompilation by up to 14.84 -- 21.58\% higher Pass@1 and Pass@10, and on binary similarity detection by up to 6.17\% Recall@1.
    \item Availability: we release \ours{} models at {\small \url{https://huggingface.co/lt-asset/nova-1.3b}} and {\small \url{https://huggingface.co/lt-asset/nova-6.7b}}

\end{itemize}

\section{Related Work}

\subsection{Binary Models}
Machine learning models are widely used in binary program analysis tasks. However, these models are typically designed for specific tasks such as binary code similarity detection~\citep{pei2020trex,xu2023diemph,jtrans, related-bcsd-1, asm2vec}, variable name prediction~\citep{dirty,xu2023lmpa,osprey, debin, dire}, binary code type inference~\citep{pei2021stateformer}, and so on~\citep{related-reverse-1, related-decompile-2, related-decompile-1}. 

Recent techniques have started to pre-train foundation LLMs for binaries. CodeArt~\citep{codeart} pre-trains encoder-style LLMs with a regularized attention design to better encode assembly code semantics. SLaDe~\citep{slade} trains BART~\citep{bart} models on assembly.
Meta LLMCompiler~\citep{llm-compiler} train CodeLlama models on LLVM IR to optimize binary code.
LLM4Decompile~\citep{llm4decompile} trains DeepSeekCoder with assembly for binary code decompilation. However, CodeArt does not generalize to generation tasks due to its encoder architecture. LLMCompiler trained on LLVM IR cannot be effectively transferred to assembly code. SLaDe and LLM4Decompile are limited in performance due to a lack of special designs for assembly. In contrast, \ours{} proposes hierarchical attention and contrastive learning objectives, outperforming existing techniques on both understanding (binary code similarity detection) and generation (binary code decompilation) tasks.

\subsection{Large Source-Code Models}
LLMs demonstrate promising results on many code-related tasks, such as code generation~\citep{related-generation-1,related-generation-2,related-generation-3,related-generation-4, codet5, codex, codegen, incoder, codellama, deepseekcoder, starcoder2, qwencoder}, bug fixing~\citep{related-apr-1,related-apr-2} and vulnerability fixing~\citep{related-vul-1, related-vul-2,related-vul-4}. The advances in using LLMs are attributed to the knowledge learned from massive source code and natural language text in their training datasets~\citep{llama,openai2023gpt4}. \ours{} is designed and trained for assembly, which has unique challenges such as low information density and diverse optimization.

\subsection{Attention Mechanism}
Standard self-attention is widely used in transformer architecture~\citep{attention} to capture soft dependencies between tokens in the input. Many special attention mechanisms have been designed for better learning in various scenarios~\citep{hierarchical-attention, long-context-survey}. LongCoder~\citep{longcoder} combines window attention and global attention (attention sink~\citep{streamingllm}) to handle long input of source code. We have shown that LongCoder's window attention is less effective than Nova's on assembly code. CAST~\citep{cast} is a new neural architecture that splits the abstract syntax tree (AST) of source code into subtrees, encodes the subtrees, and aggregates to the final encoding. PA-former~\citep{pyramid-attention} is a new neural architecture that constructs source code as pyramid input based on their AST structure and contains a pyramid attention mechanism to calculate the features in a hierarchical way. HierarchyNet~\citep{hierarchy-net} is a neural architecture that considers source code AST, data flow, and control flow graphs. Similarly, it cannot be applied to assembly code. Different from CAST, PA-former, and HierarchiyNet, Nova’s attention design is for assembly code, is more lightweight and can be plugged into any pre-trained generative LLM.

\section{Approach}
Figure~\ref{fig:overview} presents the overall approach of \ours. We build \ours{} on top of foundation models for source code~\citep{deepseekcoder} to utilize their source code and natural language generation ability. 
We first collect large assembly corpora (Section~\ref{sec:data_collection}). 
Section~\ref{sec:hierarchical_attention} describes Nova's hierarchical attention design. 
With the collected assembly corpora, we then pretrain \ours{} with language modeling and contrastive learning objectives (Section~\ref{sec:contrastive_learning}). 
Then, we fine-tune \ours{} on two important downstream tasks, binary code decompilation, and binary code similarity detection (Sections~\ref{sec:decompilation} and~\ref{sec:similarity}), to prove \ours{}'s effectiveness and benefits to the binary research domain.

\begin{minipage}[htp]{\linewidth}
    \centering
    \begin{minipage}{0.47\linewidth}
        \centering
        \includegraphics[width=\linewidth]{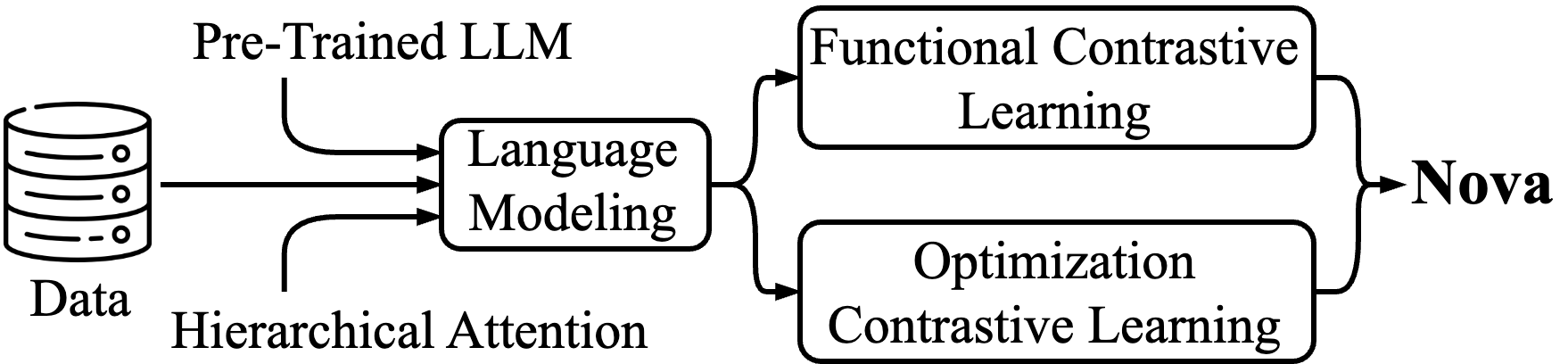}
        \captionof{figure}{Overview of developing \ours{}}
        \label{fig:overview}
    \end{minipage}
    \hspace{0.01\linewidth}
    \begin{minipage}{0.5\linewidth}
        \centering
        \scriptsize
        \setlength{\tabcolsep}{1.9pt}
        \captionof{table}{Statistics (number of \rev{C} and \rev{X86-64} assembly functions) of the pre-training datasets.}
        \begin{tabular}{lcccccl}
        \toprule
            \rev{Origin} & \rev{C Functions} & O0 & O1 & O2 & O3  & Total\\
        \midrule
            AnghaBench & 757.1K & 743.1K & 726.4K & 718.7K & 717.8K & 3.7M \\
            The-Stack & 138.8K & 125.1K & 119.7K & 116.9K & 108.8K & 609.3K \\
        \bottomrule
        \end{tabular}
        \label{tab:pretrain_dataset}
    \end{minipage}
\end{minipage}

\subsection{Data Collection}
\label{sec:data_collection}
\rev{In this paper, we focus on X86-64 assembly functions for C programs. Yet, \ours{}'s approach is generalizable to other assembly languages such as ARM assembly.}

\rev{We derive our X86-64 assembly functions dataset from two source code corpora: C functions in The-Stack~\citep{starcoder} and AnghaBench~\citep{anghabench}}.
We compile the \rev{C} programs into executables \rev{using \code{gcc}} with different optimization levels (i.e., \code{O0}, \code{O1}, \code{O2} and \code{O3}), strip the executables to remove debug information, and disassemble them into \rev{X86-64} assembly code \rev{using \code{objdump}}. We treat every function as a separate data point. \rev{Table~\ref{tab:pretrain_dataset} shows the number of C functions in the two original datasets, and the number of X86-64 assembly functions we collected from them}.

We perform certain normalization on the assembly functions: (1) removing all the ``\code{\%}'' and comments, (2) adding whitespace around ``\code{,}'', ``\code{(}'', ``\code{)}'', (3) converting all the hexadecimal numbers to decimal numbers, and (4) replacing the address of each instruction with special labels (e.g., replacing ``\code{0}'' and ``\code{4}'' in Figure~\ref{fig:motivation} (b) with ``\code{[INST-1]}'' and ``\code{[INST-2]}'') placing at the end of each instruction. More details are in Appendix~\ref{sec:appendix_data_collection}.

\subsection{Hierarchical Self-Attention}
\label{sec:hierarchical_attention}
\ours{} uses hierarchical self-attention that is specially designed to learn the \emph{low-information-density} semantics in the \emph{long} sequence of assembly code. Specifically, \ours{} learns the assembly code in an
hierarchical way by providing a modified attention mask. Different from standard token-level attentions~\citep{attention, gpt-1, gpt-2, gpt-3}, 
our hierarchical self-attention contains three categories at different levels of granularity.

\begin{figure}[htp]
    \centering
    \includegraphics[width=0.85\linewidth]{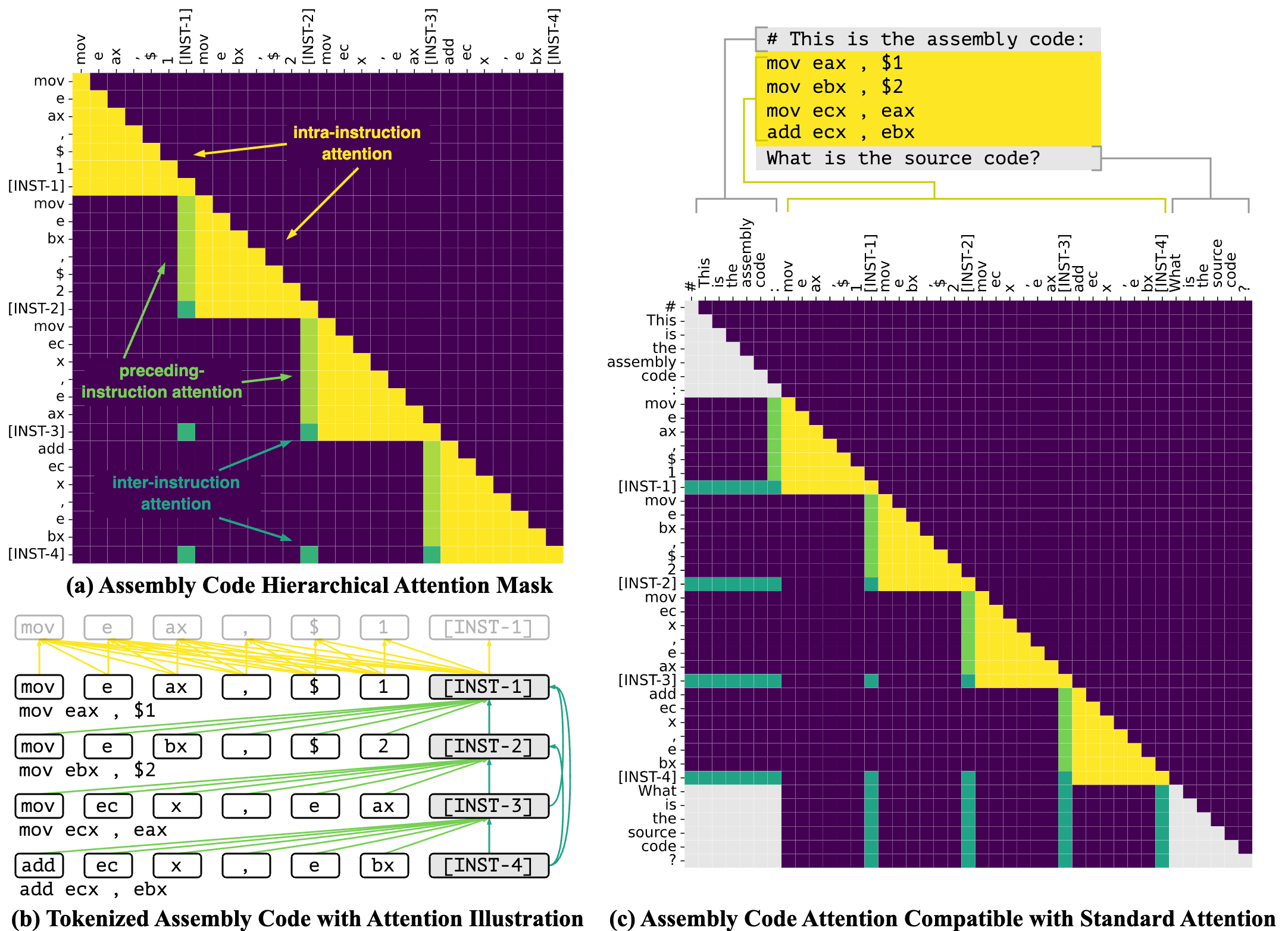}
    \caption{Design of \ours's hierarchical attention for assembly code}
    \label{fig:attention}
\end{figure}

\noindent \textbf{(1) Intra-Instruction Attention:} 
Due to the low information density in assembly, intra-instruction attention is designed to capture the summary of every instruction, which is the standard causal attention but limited to tokens of each instruction (the \colorbox{yellow}{\raisebox{0pt}[5pt][0pt]{yellow}} part in Figures~\ref{fig:attention}) (a) and (b).
Tokens in different instructions have no attention weights. The ``\code{[INST]}'' label at the end of the instruction has attention to all the tokens in the instruction and thus captures the semantics of the entire instruction (e.g., ``\code{[INST-1]}'' captures the semantics of ``\code{mov eax, \$1}''). 

\noindent \textbf{(2) Preceding-Instruction Attention:}
In addition to the local semantics of each instruction,  the use of assembly instructions (such as the choice of registers) depends on the context. 
For example, after the first instruction ``\code{mov eax, \$1}'', the second instruction should not reuse ``\code{eax}'' to store another value ``\code{\$2}'' immediately. To capture such context, the preceding-instruction attention enables each token in an instruction to have attention to the ``\code{[INST]}'' label of the preceding instruction (the \colorbox{lightgreen}{\raisebox{0pt}[5pt][0pt]{light green}} part in Figures~\ref{fig:attention} (a) and (b)).

\noindent \textbf{(3) Inter-Instruction Attention:}
To understand function semantics (i.e., functionality), which lies in the dependencies across different instructions, the inter-instruction attention is designed to let the ``\code{[INST]}'' label of each instruction have attention to all the labels of previous instructions. For example, ``\code{[INST-4]}'' has attention to ``\code{[INST-1]}'', ``\code{[INST-2]}'', and ``\code{[INST-3]}'' (the \colorbox{darkgreen}{\raisebox{0pt}[5pt][0pt]{dark green}} part in Figures~\ref{fig:attention} (a) and (b)). The inter-instruction attention is only enabled for ``\code{[INST]}'' labels, as they represent the semantics of each instruction.

To sum up, the hierarchical self-attention splits assembly code semantics into three levels: intra-instruction attention captures instruction summaries, preceding-instruction attention provides context from the preceding instruction, and inter-instruction attention models long dependencies across instructions using ``\code{[INST]}'' tokens that contain the instruction summary. Figure~\ref{fig:attention} (c) highlights the compatibility of \ours's hierarchical attention with standard self-attention for text and source code. Leveraging the proven performance of standard self-attention in existing LLMs, we retain the causal attention mask within and across chunks of text or source code (shown in light grey in Figure~\ref{fig:attention} (c)). Cross-attention between text, source code, and assembly is restricted to ``\code{[INST]}'' tokens, which encapsulate assembly instruction summaries.

\subsection{Contrastive Learning}
\label{sec:contrastive_learning}
The syntax gap between assembly code and source code, and syntax diversity between differently-optimized assembly code make LLMs struggle to distinguish the semantics behind the syntax. \ours{} adopts contrastive learning technique~\citep{simcse} during pre-training to train LLMs to encode assembly code meaningfully w.r.t semantics.
The standard pre-training objective is language modeling by minimizing the negative likelihood of code in the pre-training corpus~\citep{gpt-1}, notated as $L_{lm}$. In addition, \ours{} is pre-trained with two new objectives, $L_{fcl}$ for functionality contrastive learning and $L_{ocl}$ for optimization contrastive learning.

\begin{figure*}[htp]
    \centering
    \includegraphics[width=\linewidth]{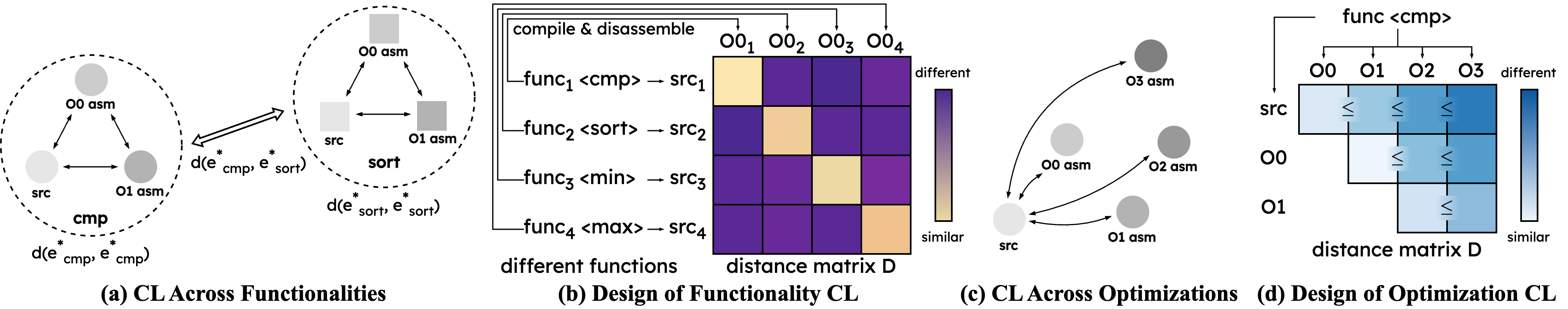}
    \caption{Design of functionality and optimization contrastive learning (CL). ``\texttt{asm}'' denotes assembly.}
    \label{fig:contrastive}
\end{figure*}

\smallskip
\noindent \textbf{Functionality CL:}
Functionality CL trains \ours{} to focus more on the functionalities of assembly code rather than the syntax. Code with the same functionality (assemblies from the same source code), should be encoded closer in the latent space. For instance, in Figure~\ref{fig:contrastive} (a), embeddings of source and assembly code of function ``\code{cmp}'' are closer to each other, and the same for function ``\code{sort}''. 

\rev{\ours{} is designed and implemented on decoder-only generative LLMs, and we refer the hidden states from the last transformer layer as embedding. For source code, we use the average of each token's embedding as the source code function's embedding. For assembly, we use the average of all the ``\code{[INST]}'' tokens' embedding as the embedding of the assembly function, as each ``\code{[INST]}'' token is supposed to capture the semantics of that instruction by the design of our hierarchical self-attention.}

Let $e^{s}_{f}$ be the embedding of function $f$ in $s$ form ($s=-1$ for source code, and $s\in[0,1,2,3]$ for \code{O0} to \code{O3} optimized assembly). For simplicity, let $S=[-1,0,1,2,3]$ be the domain of $s$. Functionality CL optimizes \rev{\ours{}'s embeddings to satisfy} the constraint:

{
\footnotesize
\begin{equation*}
    \forall f_i \in F,\; \mathop{max}\limits_{s,t \in S}(\mathop{d}(e^s_{f_i}, e^t_{f_i})) < \mathop{min}\limits_{\substack{s,t \in S \\ f_j \ne f_i \in F}}(\mathop{d}(e^s_{f_i}, e^t_{f_j}))
\end{equation*}
}
, where $\mathop{d}$ calculates the $l_2$ distance between two embeddings and $F$ is the full set of functions in the training corpus.

\rev{The embeddings of a batch of functions, each represented in two different forms, can be optimized to satisfy these constraints.}
For the example in Figure~\ref{fig:contrastive} (b), there are two forms (source code and \code{O0} assembly) of four functions. Once \ours{} encodes the batch of source code and assembly functions, we calculate the distance matrix {\small $\{D_{ij}\}_{f_i,f_j \in F} = \{d(e^s_{f_i}, e^t_{f_j})\}$}, and minimize the loss:

{
\footnotesize
\begin{equation*}
    L_{fcl} = -\mathop{log} \sum_{s,t \in S} \sum_{f_i \in F} \left(1-\frac{\mathop{exp}\left(d(e^s_{f_i}, e^t_{f_i})\right)}{\sum_{f_j \in F} \mathop{exp} \left( d(e^s_{f_i}, e^t_{f_j})\right)}\right)
\end{equation*}
}

\rev{This objective minimizes the distance between the embeddings of the same function, which is the diagonal in the distance matrix. In practice, this loss is computed and back-propagated for each batch of functions. Given a batch size of 64, $F$ represents a set of 64 unique functions in the batch.}

\smallskip
\noindent \textbf{Optimization CL:} LLMs can be confused if being asked to directly connect a source code function to its \code{O3}-optimized assembly, due to their dramatically different syntax. Such a huge gap can be filled by learning how the source code is transformed to \code{O0}, \code{O1}, \code{O2} and eventually to \code{O3} assembly, as the optimization levels are \emph{ordered}. Higher-level optimization applies a super-set of optimization rules compared to lower-level optimization. 

\ours{} learns such order with the optimization CL objective, encoding differently-optimized assembly code orderly. Optimization CL optimizes \ours{} with the constraint:
the more optimizations applied, the larger the difference between embeddings of optimized and unoptimized code. For instance, Figure~\ref{fig:contrastive} (c) and (d) illustrate that for the same function ``\code{cmp}'', the distance between source code and assembly increases when the optimization level increases. Formally, optimization CL minimizes the following loss:

{
\footnotesize
\begin{equation*}
    L_{ocl} = \sum\limits_{f \in F} \sum\limits_{s<t_1<t_2 \in S} \mathop{max}\left(0, d(e^s_{f}, e^{t_1}_{f}) - d(e^s_{f}, e^{t_2}_{f})\right)
\end{equation*}
}

Overall, the final training loss combines the three: {\small $L = L_{lm} + \lambda (L_{fcl} + L_{ocl})$}, where {\small $\lambda$} is set to 0.1 to balance the losses in this work.

\subsection{Task 1: Binary Code Decompilation}
\label{sec:decompilation}
Binary code decompilation (BCD) helps developers to understand binary code by recovering binary code into more readable high-level source code (e.g., C programs)~\citep{coda, neutron, slade, llm4decompile}. 
The input to the model for BCD is formatted as an instruction prompt (notated by $\mathbf{p}$): \code{\# This is the assembly code with \{opt\} optimization:\;\{$\mathbf{asm}$\}}, where ``\code{opt}'' is the optimization-level applied to the assembly and ``$\mathbf{asm}$'' is the assembly code to decompile. \ours{} is fine-tuned to generate the expected source code function $\mathbf{src}$ following the instruction prompt. The fine-tuning objective is minimizing the loss: {\small $L_{bcd} = -\mathop{log}\mathop{P}(\mathbf{src}|\mathbf{p})$}.

\subsection{Task 2: Binary Code Similarity Detection}
\label{sec:similarity}
Binary code similarity detection (BCSD) aims to measure the similarity between two binary code snippets~\citep{jtrans,codeart}, which is the foundation of various applications such as plagiarism detection~\citep{plagiarism-1,plagiarism-2} and vulnerability detection~\citep{binary-vul-1,binary-vul-2,binary-vul-3,binary-vul-4}.

A widely used setting is taking a query assembly of the function $f^q$ that is compiled with one optimization level (denoted by $s$), and a pool of candidate assembly of $K$ \rev{(e.g., 50, 100, etc.)} functions (notated by $f^p_i,$ {\small$1 \le i \le K$}) compiled with a different optimization level (denoted by $t \ne s$). There exists a unique candidate assembly coming from the same source code as the query ({\small $\exists! 1 \le i \le K,$} $f^p_i = f^q$, called the positive candidate). \ours{} is fine-tuned to encode these binaries, so that the positive candidate has the highest similarity with the query assembly among the pool. The learning objective is as follows:

{
\footnotesize
\begin{equation*}
    L_{BCSD} = -\mathop{log}\sum\limits_{1 \le i \le K}^{f^q:=f^p_i} \left(1-\frac{\mathop{exp}\left(\mathop{d}(e^s_{f^q}, e^t_{f^p_i})\right)}{\sum_{1 \le j \le K} \mathop{exp}\Big(\mathop{d}(e^s_{f^q}, e^t_{f^p_j})\Big)}\right)
\end{equation*}
}

We follow previous work~\citep{codeart} to let $s$ be \code{O0}-assembly and $t$ be \code{O3}-assembly, which is the hardest setting.

\section{Experimental Setup}
This section describes the setup of pre-training and fine-tuning of \ours{}, as well as the existing baselines we compare \ours{} with, and the evaluation metrics we used in the two downstream tasks.
Appendix~\ref{sec:appendix_training} contains additional details such as training hyper-parameters, and evaluation setup.

\subsection{Pre-Training}
\label{sec:pretrain_setup}
We use the \rev{C and X86-64 assembly functions} collected from AnghaBench and The-Stack for pre-training. We pre-train \ours{} starting from DeepSeek-Coder~\citep{deepseekcoder}, and the hierarchical attention is applied on half of the attention heads to balance between its effectiveness and the existing knowledge in the standard attention layers (Justified in Appendix~\ref{sec:appendix_half}). \ours{} is pre-trained with language modeling for one epoch, followed by contrastive learning objectives for another epoch. 

\subsection{Fine-Tuning for Binary Code Decompilation}
\noindent \textbf{Training Data:} We sample (due to computation resource limitation) 2.16M assembly-to-source-code pairs (0.338B tokens) from the pre-training corpus to build the BCD fine-tuning data.

\noindent \textbf{Test Data:} 
We use HumanEval-Decompile~\citep{llm4decompile} as the test benchmark, which was not used in training.
HumanEval-Decompile is derived from the C language adaptation of the HumanEval~\citep{codex} benchmark and contains 164 C functions, each compiled with \code{O0} -- \code{O3} optimization flags and disassembled into X86-64 assembly. 

\noindent \textbf{Baselines:} \ours{} is compared with 
existing SOTA LLM4Decompile~\citep{llm4decompile}, Meta LLMCompiler~\citep{llm-compiler}, other open-sourced general code LLMs~\citep{codellama, starcoder2, deepseekcoder, qwencoder, granite, longcoder}, and commercial LLMs GPT-3.5-Turbo and GPT-4o). LLM4Decompile trains DeepSeekCoder using the same AnghaBench corpus for binary decompilation. Meta LLMCompiler trains CodeLlama models using LLVM IRs, X86, and ARM assembly code to optimize and translate binary code.

\noindent \textbf{Evaluation:} 
\rev{We let each model sample 20 decompilations per assembly function, using the temperature of 0.2 and \texttt{top\_p} of 0.95~\citep{codex}. Except for LLM4Decompile and \ours{} that are fine-tuned for binary code decompilation, we provide all other baselines with three-shot examples for few-shot learning~\citep{gpt-3}. The decompilations are executed with the test cases to verify the functional correctness. Finally, Pass@1 and Pass@10~\citep{codex} are reported.}

\subsection{Fine-Tuning for Binary Code Similarity Detection}
\noindent \textbf{Training Data:} To compare \ours{} with existing works on BCSD fairly~\citep{jtrans, codeart}, we use the same dataset, BinaryCorp-3M~\citep{jtrans}, as the fine-tuning data for BCSD, which contains the \code{O0} and \code{O3} assembly of 224,606 functions.

\noindent \textbf{Test Data:} Following existing work~\citep{codeart, xu2023diemph}, we use real-world benchmarks, Binutils, Curl, ImageMagick, SQLite, OpenSSL, and Putty, as the test benchmarks, which are nonexistent in the training data.

\noindent \textbf{Baselines:} \ours{} is compared with jTrans~\citep{jtrans}, DiEmph~\citep{xu2023diemph} and CodeArt~\citep{codeart}. jTrans is a Transformer~\citep{attention} encoder trained on binaries with masked token prediction and jump target prediction tasks. DiEmph uses an instruction deemphasis technique to prevent the model from learning instruction
distribution biases introduced by compilers. CodeArt proposes a regularized attention mask for encoder models to capture instructional semantics and data dependencies.

\noindent \textbf{Evaluation:} We randomly sample \rev{$K=$ 50, 100, 200, 500} source code functions from each project, compile them into binaries with \code{O0} and \code{O3} optimization flags, and disassemble them into X86-64 assemblies. BCSD techniques encode these assemblies into embeddings (\ours{} uses the average last-layer hidden states of all the ``\code{[INST]}'' tokens in an assembly as its embedding). Then each \code{O0} assembly is used as the query to calculate their similarity with the $K$ \code{O3} candidate assemblies. Metric Recall@1 is reported as the ratio of queries for which the candidate from the same source code has the highest similarity among all the candidates.

\section{Results}

\subsection{Binary Code Decompilation}
\textbf{Comparison with SOTA Techniques:}
Table~\ref{tab:bcd} shows the Pass@1 and Pass@10 of the decompiled code from assemblies on HumanEval-Decompile. The results are grouped by optimization level (i.e., the benchmark contains 164 assemblies of each optimization level), and the average is also reported.

\begin{table}[htp]
    \scriptsize
    \centering
    \caption{\ours's Pass@K and comparison with existing techniques on HumanEval-Decompile.}
    \begin{tabular}{l|rrrrr|rrrrr}
    \toprule
        \multirow{2}{*}{Techniques} & \multicolumn{5}{c|}{Pass@1} & \multicolumn{5}{c}{Pass@10} \\
         & O0 & O1 & O2 & O3 & Avg. & O0 & O1 & O2 & O3 & Avg. \\
    \midrule
        CodeLlama-7B & 6.95 & 3.81 & 4.54 & 3.78 & 4.77 & 8.53 & 5.97 & 7.34 & 5.17 & 6.75 \\
        StarCoder2-7B & 6.31 & 4.33 & 5.64 & 5.95 & 5.56 & 8.77 & 5.18 & 6.09 & 7.17 & 6.80 \\
        DeepSeekCoder-7B & 9.63 & 7.56 & 7.41 & 6.68 & 7.82 & 13.60 & 11.38 & 11.52 & 9.44 & 11.49 \\
        Qwen-2.5-Coder-7B & 4.76& 5.79& 5.58& 5.27& 5.35& 6.34& 7.69& 6.56& 5.79& 6.60\\
        LLMCompiler-7B & 5.95 & 5.85 & 5.55 & 5.82 & 5.79 & 7.01 & 7.31 & 7.47 & 7.01 & 7.20 \\
        GPT-3.5-Turbo & 7.41 & 6.13 & 4.33 & 3.90 & 5.44 & 9.56 & 8.38 & 6.23 & 5.12 & 7.32 \\
        GPT-4o & 21.34 & 18.29 & 14.48 & 13.05 & 16.79 & 29.94 & 26.74 & 21.42 & 19.88 & 24.50 \\
    \midrule
        LLM4Decompile-1.3B & 15.30 & 8.26 & 9.36 & 8.38 & 10.33 & 21.79 & 15.23 & 16.17 & 13.70 & 16.72 \\
        \textbf{Nova-1.3B} & \textbf{37.53}& \textbf{21.71}& \textbf{22.68}& \textbf{18.75}& \textbf{25.17}& \textbf{49.38}& \textbf{34.84}& \textbf{36.95}& \textbf{32.03}& \textbf{38.30}\\
    \midrule
        LLM4Decompile-6.7B & 29.97 & 19.05 & 20.46 & 18.32 & 21.95 & 40.40 & 27.75 & 28.85 & 28.51 & 31.38 \\
        \textbf{Nova-6.7B} & \textbf{48.78} & \textbf{30.58} & \textbf{30.85} & \textbf{27.23} & \textbf{34.36} & \textbf{57.47} & \textbf{47.45} &\textbf{ 43.03} & \textbf{39.68} & \textbf{46.91} \\
    \bottomrule
    \end{tabular}
    \label{tab:bcd}
\end{table}

\emph{Overall, \ours{}'s Pass@1 and Pass@10 are higher than all SOTA binary decompilation techniques and general LLMs with even smaller model sizes}. 
Specifically, for each optimization level, \ours{} consistently decompiles more assemblies into source code correctly than the rest of the compared techniques. Note that Meta LLMCompiler is mainly designed for LLVM IR optimization, and thus is still incapable of assembly code decompilation.

With the same model size, \ours{}-1.3B outperforms LLM4Decompile-1.3B with a 14.84\% higher averaged Pass@1, and a 21.58\% higher Pass@10. \ours{}-6.7B outperforms LLM4Decompile-6.7B with a 12.41\% higher averaged Pass@1, and a 15.53\% higher Pass@10. When compared with GPT-4o, an order of magnitude larger model, \ours{}-1.3B produces an 8.38\% higher Pass@1 and 13.80\% higher Pass@10. 
Examples of \ours{}'s correct decompilation are provided in Appendix~\ref{sec:appendix_bcd_case_study}.

\textbf{Comparison with Techniques Handling Long Input:}
\ours{}'s hierarchical attention design targets to address the low information density and long input challenge of assembly code. There are other techniques that handles long input challenges in text and source code, with Granite-3B-Code-Base-128K and LongCoder being the most related ones.

Granite trains LLM on repository-level long inputs, which is an orthogonal approach with \ours{}'s approach (hierarchical attention and contrastive learning). We train Granite-3B-Code-128K with \ours{}'s approach, and Table~\ref{tab:bcd_granite} shows that \ours{}'s approach brings improvement to Granite over standard fine-tuning even if it has already been trained with long code data.

\begin{table}[htp]
    \scriptsize
    \centering
    \caption{\ours{}'s approach brings improvement to LLM that has been trained with long input data.}
    \begin{tabular}{l|rrrrr|rrrrr}
    \toprule
        \multirow{2}{*}{Techniques} & \multicolumn{5}{c|}{Pass@1} & \multicolumn{5}{c}{Pass@10} \\
         & O0 & O1 & O2 & O3 & Avg. & O0 & O1 & O2 & O3 & Avg. \\
    \midrule
        Granite (3B-Code-128K) & 5.91 & 3.78 & 5.09 & 5.52 & 5.08 & 8.19 & 5.16 & 6.56 & 7.15 & 6.76 \\
        Granite + Standard Fine-Tuning & 20.88 &	13.54 & 11.37 & 10.09 & 13.97 & 30.05 & 19.77 & 18.31 & 15.77 & 20.98 \\
        \textbf{Granite + \ours{}'s \rev{Approaches}} & \textbf{31.04} & \textbf{14.57} & \textbf{14.70} & \textbf{13.66} & \textbf{18.49} & \textbf{39.57} & \textbf{21.23} & \textbf{21.77} & \textbf{19.82} & \textbf{25.60} \\
    \bottomrule
    \end{tabular}
    \label{tab:bcd_granite}
\end{table}

\begin{table}[htp]
    \scriptsize
    \centering
    \setlength{\tabcolsep}{5pt}
    \caption{\ours{}'s hierarchical attention is more effective on assembly code.}
    \begin{tabular}{l|rrrrr|rrrrr}
    \toprule
        \multirow{2}{*}{Techniques} & \multicolumn{5}{c|}{Pass@1} & \multicolumn{5}{c}{Pass@10} \\
         & O0 & O1 & O2 & O3 & Avg. & O0 & O1 & O2 & O3 & Avg. \\
    \midrule
        \rev{\ours{}-1.3B (using LongCoder's Attention)} & 34.59 & 19.07 & 19.72 & 17.34 & 22.68 & 42.19 & 32.37 & 32.86 & 29.04 & 34.12 \\
        \textbf{\rev{\ours{}-1.3B}} & \textbf{37.53} & \textbf{21.71} & \textbf{22.68} & \textbf{18.75} & \textbf{25.17} & \textbf{49.38} & \textbf{34.84} & \textbf{36.95} & \textbf{32.03} & \textbf{38.30} \\
    \bottomrule
    \end{tabular}
    \label{tab:bcd_longcoder}
\end{table}

LongCoder combines window attention and global attention to learn long code input. We compare LongCoder's attention design with \ours{}'s hierarchical attention design \rev{by replacing the hierarchical attention of \ours{}-1.3B with LongCoder's attention design}. Table~\ref{tab:bcd_longcoder} shows that \ours{}'s hierarchical attention is more effective in learning assembly code. Nova’s attention design considers the instruction-level local semantics and dependencies between different instructions, which fits better than fix-sized window attention to assembly code.

\textbf{Ablation Study:}
We conduct an ablation study by comparing \ours{}-1.3B with the following models to show the effectiveness of contrastive learning objectives and hierarchical attention design:
\begin{itemize}[leftmargin=15pt, noitemsep]
    \item \nocl: Removing contrastive learning and hierarchical self-attention. This is simply training DeepSeekCoder-1.3B on the assembly corpus using language modeling. This can be viewed as our reproduction (retrain) of LLM4Decompile-1.3B using the same amount of data.
    \item \noha: Removing the hierarchical self-attention, training DeepSeekCoder-1.3B on the assembly corpus using both the language modeling and contrastive learning objectives.
\end{itemize}

\begin{table}[t]
    \scriptsize
    \centering
    \caption{Ablation study of Nova-1.3B on HumanEval-Decompile.}
    \begin{tabular}{l|rrrrr|rrrrr}
    \toprule
        \multirow{2}{*}{Techniques} & \multicolumn{5}{c|}{Pass@1} & \multicolumn{5}{c}{Pass@10} \\
         & O0 & O1 & O2 & O3 & Avg. & O0 & O1 & O2 & O3 & Avg. \\
    \midrule
        LLM4Decompile-1.3B & 15.30 & 8.26 & 9.36 & 8.38 & 10.33 & 21.79 & 15.23 & 16.17 & 13.70 & 16.72 \\
    \midrule
        \nocl & 20.73& 16.16& 15.03& 11.19& 15.78& 33.55& 28.12& 26.96& 21.01& 27.41\\
        \noha & 30.58& 19.88& 20.58& 16.40& 21.86& 44.75& 33.13& 33.31& 29.82& 35.25\\
        \textbf{Nova} & \textbf{37.53}& \textbf{21.71}& \textbf{22.68}& \textbf{18.75}& \textbf{25.17}& \textbf{49.38}& \textbf{34.84}& \textbf{36.95}& \textbf{32.03}& \textbf{38.30}\\
    \bottomrule
    \end{tabular}
    \label{tab:bcd_ablation}
\end{table}

Table~\ref{tab:bcd_ablation} shows the results of the ablation study. \nocl{} produces an average Pass@1 of 15.78\% and Pass@10 of 27.41\%. With additional contrastive learning objectives, \noha{} improves the Pass@1 on all optimization levels over \nocl{}, showing a higher averaged Pass@1 and Pass@10. Further applying the hierarchical self-attention on \noha{} boosts the overall Pass@1 from 21.86\% to 25.17\%, and Pass@10 from 35.25\% to 38.30\%. 

\begin{table*}[htp]
    \scriptsize
    \centering
    \setlength{\tabcolsep}{3pt}
    \begin{tabular}{cc}
        \begin{minipage}{0.49\textwidth}
            \centering
            \setlength{\tabcolsep}{2.2pt}
            \caption{Recall@1 on BCSD with {\small $K=50$}}
            \begin{tabular}{l|ccc>{\columncolor[gray]{0.9}}c|>{\columncolor[gray]{0.85}}c}
                \toprule
                    Benchmarks & jTrans & DiEmph & CodeArt & \ours{}-1.3B & \ours{}-6.7B \\
                \midrule
                    Binutils & 0.68 & 0.80 & 0.84 & \underline{0.87}& 0.89\\
                    Curl & 0.72 & 0.84 & 0.86 & \underline{0.89} & 0.94 \\
                    ImageMagick & 0.53 & 0.71 & 0.78 & \underline{0.86} & 0.90 \\
                    SQLite & 0.73 & \underline{0.79} & 0.78 & 0.77 & 0.78 \\
                    OpenSSL & 0.70 & 0.83 & 0.88 & \underline{0.90} & 0.92 \\
                    Putty & 0.63 & \uwave{0.72} & 0.69 & \uwave{0.72} & 0.71 \\
                \midrule
                    Avg. & 0.67 & 0.78 & 0.81 & \textbf{0.84} & \textbf{0.86} \\
                \bottomrule
            \end{tabular}
            \label{tab:bcsd_k_50}
        \end{minipage} &
        \begin{minipage}{0.49\textwidth}
            \centering
            \setlength{\tabcolsep}{2.2pt}
            \caption{Recall@1 on BCSD with {\small $K=100$}}
            \begin{tabular}{l|ccc>{\columncolor[gray]{0.9}}c|>{\columncolor[gray]{0.85}}c}
                \toprule
                    Benchmarks & jTrans & DiEmph & CodeArt & \ours{}-1.3B & \ours{}-6.7B \\
                \midrule
                    Binutils & 0.60 & 0.63 & \underline{0.81} & 0.79 & 0.79 \\
                    Curl & 0.63 & 0.80 & 0.82 & \underline{0.86} & 0.88 \\
                    ImageMagick & 0.54 & 0.71 & 0.76 & \underline{0.79} & 0.81 \\
                    SQLite & 0.62 & 0.72 & \underline{0.74} & 0.73 & 0.72 \\
                    OpenSSL & 0.60 & 0.80 & 0.87 & \underline{0.88} & 0.90 \\
                    Putty & 0.58 & 0.64 & 0.64 & \underline{0.65} & 0.64 \\
                \midrule
                    Avg. & 0.60 & 0.72 & 0.77 & \textbf{0.78} & \textbf{0.79} \\
                \bottomrule
            \end{tabular}
            \label{tab:bcsd_k_100}
        \end{minipage} \\
        \multicolumn{2}{c}{} \\[0.1em]
        \begin{minipage}{0.49\textwidth}
            \centering
            \setlength{\tabcolsep}{2.2pt}
            \caption{Recall@1 on BCSD with {\small $K=200$}}
            \begin{tabular}{l|ccc>{\columncolor[gray]{0.9}}c|>{\columncolor[gray]{0.85}}c}
                \toprule
                    Benchmarks & jTrans & DiEmph & CodeArt & \ours{}-1.3B & \ours{}-6.7B \\
                \midrule
                    Binutils & 0.51 & 0.64 & \underline{0.74} & 0.73 & 0.73 \\
                    Curl & 0.57 & 0.77 & 0.78 & \underline{0.83} & 0.84 \\
                    ImageMagick & 0.39 & 0.51 & 0.67 & \underline{0.73} & 0.75 \\
                    SQLite & 0.56 & 0.65 & \uwave{0.68} & \uwave{0.68} & 0.69 \\
                    OpenSSL & 0.54 & 0.71 & 0.82 & \underline{0.84} & 0.88 \\
                    Putty & 0.49 & \underline{0.58} & 0.55 & 0.55 & 0.58 \\
                \midrule
                    Avg. & 0.51 & 0.64 & 0.71 & \textbf{0.73} & \textbf{0.75} \\
                \bottomrule
            \end{tabular}
            \label{tab:bcsd_k_200}
        \end{minipage} &
        \begin{minipage}{0.49\textwidth}
            \centering
            \setlength{\tabcolsep}{2.2pt}
            \caption{Recall@1 on BCSD with {\small $K=500$}}
            \begin{tabular}{l|ccc>{\columncolor[gray]{0.9}}c|>{\columncolor[gray]{0.85}}c}
                \toprule
                    Benchmarks & jTrans & DiEmph & CodeArt & \ours{}-1.3B & \ours{}-6.7B \\
                \midrule
                    Binutils & 0.40 & 0.57 & \underline{0.70} & 0.65 & 0.67 \\
                    Curl & 0.43 & 0.62 & 0.69 & \underline{0.73} & 0.76 \\
                    ImageMagick & 0.25 & 0.42 & 0.58 & \underline{0.61} & 0.65 \\
                    SQLite & 0.43 & 0.59 & \underline{0.62} & 0.59 & 0.62 \\
                    OpenSSL & 0.43 & 0.61 & 0.76 & \underline{0.78} & 0.82 \\
                    Putty & 0.38 & \underline{0.50} & 0.49 & 0.47 & 0.51 \\
                \midrule
                    Avg. & 0.39 & 0.55 & \textbf{0.64} & \textbf{0.64} & \textbf{0.67} \\
                \bottomrule
            \end{tabular}
            \label{tab:bcsd_k_500}
        \end{minipage} \\
    \end{tabular}
\end{table*}

\subsection{Binary Code Similarity Detection}
Tables~\ref{tab:bcsd_k_50}, ~\ref{tab:bcsd_k_100}, ~\ref{tab:bcsd_k_200} and ~\ref{tab:bcsd_k_500} show the Recall@1 of \ours{} and existing BCSD techniques with pool size $K$ of 50, 100, 200 and 500 on the six benchmarks. 
\underline{Underline} indicates the best in each benchmark, while \uwave{wave} denotes the tied best (we only mark \ours{}-1.3B for clearer illustration). 

Overall, Tables~\ref{tab:bcsd_k_50}, ~\ref{tab:bcsd_k_100}, ~\ref{tab:bcsd_k_200} and~\ref{tab:bcsd_k_500} show that \emph{on average, \ours{}-1.3B and \ours{}-6.7B  achieve the highest Recall@1 (in \textbf{bold}) under all four settings of $K$}. 
\ours{}-6.7B further outperforms \ours{}-1.3B  and achieves the highest averaged Recall@1 under all four settings, ranking the ground-truth of 5\%, 2\%, 4\%, and 3\% more queries the most similar correspondingly compared to CodeArt.
\ours{}-1.3B consistently outperforms existing techniques with higher Recall@1 when $K$ is 50, 100, and 200, meaning it correctly ranks ground-truth of 3\%, 1\%, and 2\% more queries as the most similar. Under the setting of $K=500$, \ours{}-1.3B ties with CodeArt with the same highest Recall@1. When looking into each individual benchmark, \ours{}-1.3B always wins on the most benchmarks under different settings of pool size $K$. For instance, \ours{}-1.3B wins on four benchmarks while DiEmph only wins on SQLite when $K=50$.
We also conduct an ablation study on BCSD in Appendix~\ref{sec:appendix_bcsd_ablation}.

\subsection{Analytic Experiments: How are \ours{}'s embeddings better?}
\begin{figure}[htp]
    \centering
    \includegraphics[width=0.82\linewidth]{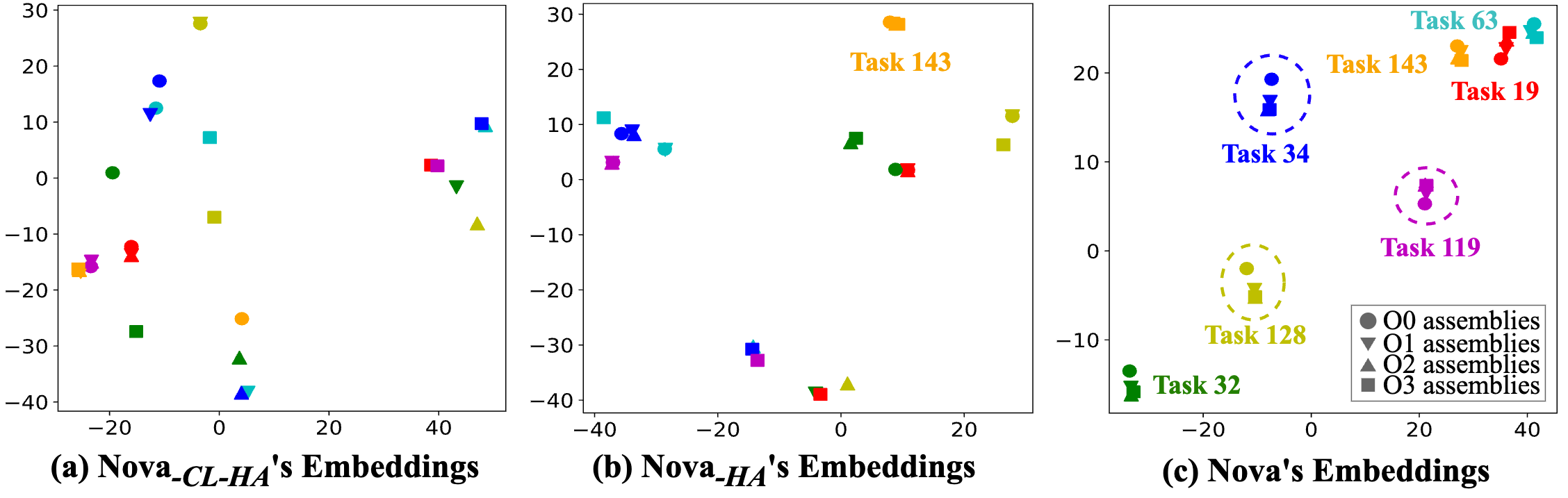}
    \caption{t-SNE analysis of embeddings calculated by \nocl{}, \noha{}, and \ours{}.}
    \label{fig:embedding_tsne}
\end{figure}

We use the widely-used t-SNE~\citep{tsne} to analyze and visualize high-dimensional embeddings. We randomly sample seven coding problems from HumenEval-Decompile (\texttt{task\_id} 19, 32, 34, 63, 119, 128, 143), encode the \code{O0} -- \code{O3}  assemblies by \nocl{} and \ours{}-1.3B. Figure~\ref{fig:embedding_tsne} shows the embeddings that are visualized under the first two principal components. Each color represents one task, and \code{O0} -- \code{O3} assemblies are marked by \scalebox{0.9}{$\bigcirc$}, $\bigtriangledown$, $\bigtriangleup$, and $\square$.

Compared with \nocl{} (Figure~\ref{fig:embedding_tsne} (a)), \noha{} (Figure~\ref{fig:embedding_tsne} (b)) including contrastive learning objectives in the pre-training, can separate the embeddings of assemblies with different functionalities better. \noha{} clearly encode ``\texttt{Task 143}'' (orange points) away from the others. Yet, \ours{}'s (Figure~\ref{fig:embedding_tsne} (c)) embeddings group the assemblies by functionalities more precisely than \noha{}, suggesting that hierarchical attention enhances the training of contrastive learning objectives to learn more effective encoding.
Visualization using a different approach, PCA, is shown in Appendix~\ref{sec:appendix_embedding_analysis}. Analytic experiments on \ours{}'s hierarchical attention is shown in Appendix~\ref{sec:appendix_attention_analysis}.

\section{Conclusion}
This work presents \ours{}, a generative foundation LLM for assembly code that introduces two key innovations—hierarchical attention and contrastive learning—to tackle the unique challenges of assembly code understanding. \ours{} builds upon a source code LLM and undergoes further pre-training on approximately 4.3 million collected assembly functions. We fine-tune and evaluate \ours{} on two downstream tasks: binary code decompilation and binary code similarity detection. \ours{} achieves up to 14.84 – 21.58\% (absolute
percentage point improvement) higher Pass@1 and Pass@10 over existing methods in binary code decompilation and up to a 6.17\% higher Recall@1 in binary code similarity detection. Looking forward, we believe that the proposed hierarchical attention and contrastive learning methods hold promise for enhancing foundation models in both source code and natural language domains, which we leave as future work.

\section*{ACKNOWLEDGMENTS}
This research was supported in part by NSF 1901242 and 2006688,  a CFI fund, and J.P. Morgan AI Faculty Research Awards. 
This work also used Anvil at Purdue University through allocation CIS240304 from the Advanced Cyberinfrastructure Coordination Ecosystem: Services \& Support (ACCESS) program~\citep{anvil}, which is supported by NSF grants 2138259, 2138286, 2138307, 2137603, and 2138296.
Any opinions, findings, and conclusions in this paper are those of the authors only and do not necessarily reflect the views of our sponsors.


\bibliography{paper}
\bibliographystyle{iclr2025_conference}

\appendix

\section{Appendix}

\subsection{Data Collection}
\label{sec:appendix_data_collection}
This section provides additional details of the data collection. To collect assemblies from The-Stack, we attempt to compile 4 million C programs, of which 138.8K is compiled successfully. We do not collect more due to the computation resource limitations. 

For the 757.1K and 138.8K source code that successfully compiled into executables (using \texttt{gcc}) from AnghaBench and The-Stack, we disassemble them using \texttt{objdump}. \texttt{objdump} was not able to successfully disassemble all the executables, resulting in some empty assembly code. Thus, the number of \code{O0} -- \code{O1} we obtain from each corpus is different and smaller than the number of source codes as shown in Table~\ref{tab:pretrain_dataset}.

\rev{There may be alternative ways of collecting assembly code, e.g., using \code{gcc -S} to directly dump the assembly code without producing executables. However, a key difference is that the assembly generated by \code{gcc -S} does not undergo the linking step. In practice, binary decompilation and analysis are typically performed on executable or linked assembly code, as it includes linker modifications and reflects the final binary structure. Our data collection aligns better with the practical use of assembly. Nevertheless, \ours{}'s approaches should be generalizable to assembly code obtained with a different approach, and the  experiments remain as future work.}

\begin{figure}[htp]
    \centering
    \includegraphics[width=0.7\linewidth]{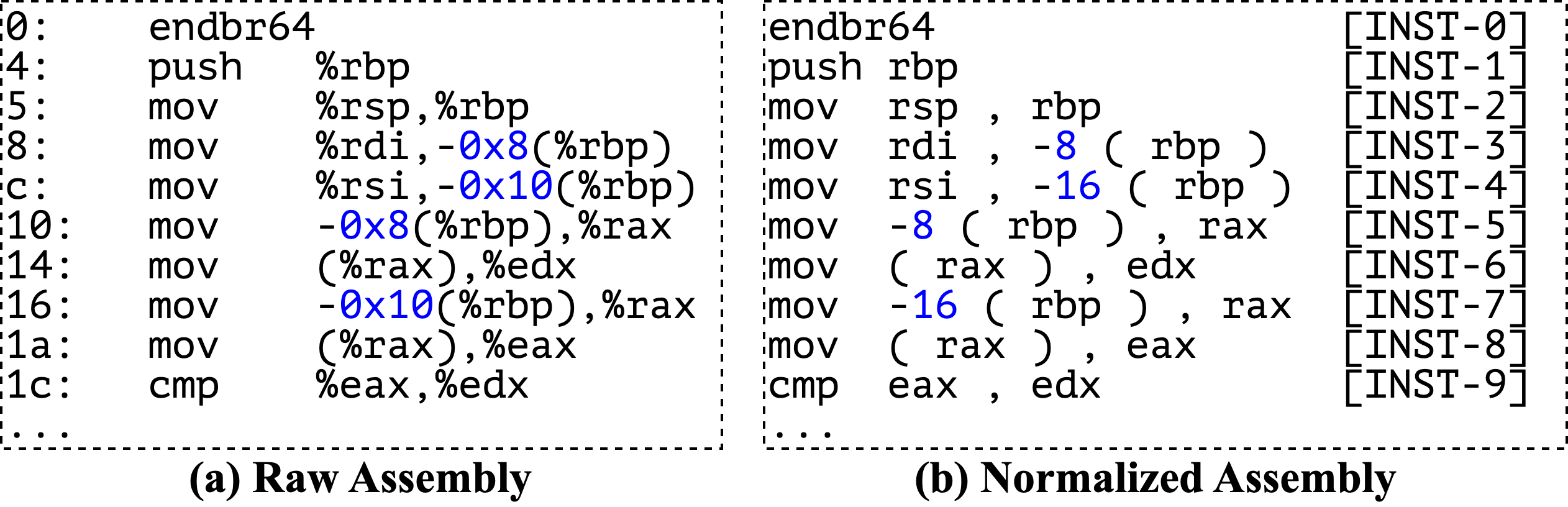}
    \caption{Example of assembly code preprocessing}
    \label{fig:appendix_normalize}
\end{figure}

Figure~\ref{fig:appendix_normalize} shows an example of preprocessing the raw assembly code as described in Section~\ref{sec:data_collection}.

\subsection{Experimental Setup Details}
\label{sec:appendix_training}
This section provides additional details of training. We pre-train \ours{} starting from DeepSeek-Coder, using the language modeling objective ({\small $L_{lm}$}) for one epoch on the \rev{C functions and assembly functions collected from} AnghaBench and The-Stack corpora. This step uses a batch size of 128, with the input truncated by a 1,024 tokens limit. The model weights are updated using the AdamW optimizer. The learning rate is $5e^{-5}$, using 1000 steps of warm-up and a cosine decay to adjust the learning rate.

Then, the model is further pre-trained with the combination of language modeling and contrastive learning objectives ({\small $L = L_{lm} + \lambda (L_{fcl} + L_{ocl})$}), with $\lambda$ set to 0.1. To train with the functionality contrastive learning objective, we discard any source code that misses any one of \code{O0} -- \code{O3} assemblies and also discard the source code whose \code{O2} assembly is the same as its \code{O3} assembly. As a result, this step is only trained for 0.36M data samples for one epoch. The batch size is 64, with the input truncated by a 1,024 tokens limit. The learning rate is $2e^{-5}$ using the AdamW optimizer. 

The fine-tuning of both BCD and BCSD uses a batch size of 64, with the input truncated by a 2,048 token limit. Similarly, the learning rate is $2e^{-5}$ using the AdamW optimizer, and the model is fine-tuned for one epoch. During the training using the contrastive learning objectives, and the fine-tuning of BCSD, we use the average of \code{[INST]} tokens’ last layer hidden states to represent the embedding of a binary function.

\rev{For the evaluation of BCSD, we reuse the framework provided by CodeArt~\cite{codeart} to evaluate the binary code similarity detection results once \ours{} produces the embeddings for functions in the test dataset. For each one of the $K$ functions, the \code{O0} assembly function is used as the query to calculate the cosine similarity between its embedding and the embeddings of all the $K$ \code{O3} assembly functions. Note that \ours{} does not normalize the embeddings of assembly functions during training and uses $l_2$ distance to calculate the $L_{BCSD}$, which optimize \ours{} to embed the \code{O0} and \code{O3} assembly functions from the same source code have the smallest $l_2$ distances. When using CodeArt's framework for evaluation which ranks the similarity using the cosine similarity, we normalize \ours{}'s embedding for each assembly function since $l_2$ distance after normalization keeps the same order as cosine similarity (smaller $l_2$ distance means higher cosine similarity).}

\smallskip
\noindent \textbf{Infrastructure} The experiments are conducted on eight NVIDIA RTX A100 GPUs, each with 40GB memory. Our implementation is based on Huggingface's implementation of DeepSeek-Coder~\footnote{\url{https://huggingface.co/deepseek-ai/deepseek-coder-1.3b-base}}, PyTorch~\footnote{\url{https://pytorch.org/get-started/locally/}}, and DeepSpeed~\footnote{\url{https://github.com/microsoft/DeepSpeed}}.

\begin{table}[t]
    \scriptsize
    \centering
    \caption{Comparison with applying hierarchical attention on all attention heads, using 1B models.}
    \begin{tabular}{l|rrrrr|rrrrr}
    \toprule
        \multirow{2}{*}{Techniques} & \multicolumn{5}{c|}{Pass@1} & \multicolumn{5}{c}{Pass@10} \\
         & O0 & O1 & O2 & O3 & Avg. & O0 & O1 & O2 & O3 & Avg. \\
    \midrule
        \noha & 30.58& 19.88& 20.58& 16.40& 21.86& 44.75& 33.13& 33.31& 29.82& 35.25 \\
        Nova (hierarchical on all heads) & 32.38 & 18.87 & 20.56 & 16.34 & 22.04 & 45.95 & 32.19 & 32.78 & 29.01 & 34.98 \\
        \textbf{\ours{}} & \textbf{37.53}& \textbf{21.71}& \textbf{22.68}& \textbf{18.75}& \textbf{25.17}& \textbf{49.38}& \textbf{34.84}& \textbf{36.95}& \textbf{32.03}& \textbf{38.30}\\
    \bottomrule
    \end{tabular}
    \label{tab:attention_head}
\end{table}

\subsection{Applying Hierarchical Attention on Half Attention Heads}
\label{sec:appendix_half}
The hierarchical attention mask is always applied on half of the attention heads at each layer in \ours{}. This ensures the LLM balances the hierarchical knowledge of assembly code and pre-trained knowledge learned by full self-attention.

We conducted experiments applying hierarchical attention to all the attention heads. Results in Table~\ref{tab:attention_head} show that 
when applying hierarchical attention to all the attention heads of transformer layers, the performance does not improve enough and even drops under some settings. This implies that the standard full self-attention mechanism indeed learns knowledge that may not be captured by hierarchical attention. Thus, we only apply hierarchical attention to half of the attention heads in each transformer layer of \ours{} to balance the knowledge learned by standard and hierarchical attention.

\begin{table}[htp]
    \scriptsize
    \centering
    \caption{Ablation study of Nova-1.3B on HumanEval-Decompile.}
    \begin{tabular}{l|rrrrr|rrrrr}
    \toprule
        \multirow{2}{*}{Techniques} & \multicolumn{5}{c|}{Pass@1} & \multicolumn{5}{c}{Pass@10} \\
         & O0 & O1 & O2 & O3 & Avg. & O0 & O1 & O2 & O3 & Avg. \\
    \midrule
        \nocl & 20.73& 16.16& 15.03& 11.19& 15.78& 33.55& 28.12& 26.96& 21.01& 27.41\\
        \rev{\nofcl} & 22.38 & 16.20 & 16.37 & 13.25 & 17.05& 36.13& 29.48& 30.02& 23.76& 29.85\\
        \rev{\noocl} & 28.44& 18.87& 18.53& 15.76& 20.40& 40.28& 32.33& 31.80& 27.05& 32.87\\
        \noha & 30.58& 19.88& 20.58& 16.40& 21.86& 44.75& 33.13& 33.31& 29.82& 35.25\\
        \textbf{Nova} & \textbf{37.53}& \textbf{21.71}& \textbf{22.68}& \textbf{18.75}& \textbf{25.17}& \textbf{49.38}& \textbf{34.84}& \textbf{36.95}& \textbf{32.03}& \textbf{38.30}\\
    \bottomrule
    \end{tabular}
    \label{tab:appendix_bcd_ablation}
\end{table}

\subsection{Additional Ablation Study on Binary Code Decompilation}
\label{sec:appendix_bcd_ablation_study}
\rev{
We provide additional ablation studies on studying the impact of each individual contrastive learning objective. We study two more models:}
\begin{itemize}[leftmargin=15pt, noitemsep]
    \item \rev{\nofcl: Removing functional contrastive learning and hierarchical self-attention.}
    \item \rev{\noocl: Removing optimization contrastive learning and hierarchical self-attention.}
\end{itemize}

\rev{
Table~\ref{tab:appendix_bcd_ablation} shows that each component of the contrastive learning brings certain improvements to the Pass@1 and Pass@10 on HumanEval-Decompile, and we find the impact of function contrastive learning (FCL) is larger than the impact of optimization contrastive learning (OCL), suggesting that aligning the model's embeddings for assembly code with the same functionality is more useful.
}

\begin{figure}[htp]
    \centering
    \includegraphics[width=0.75\linewidth]{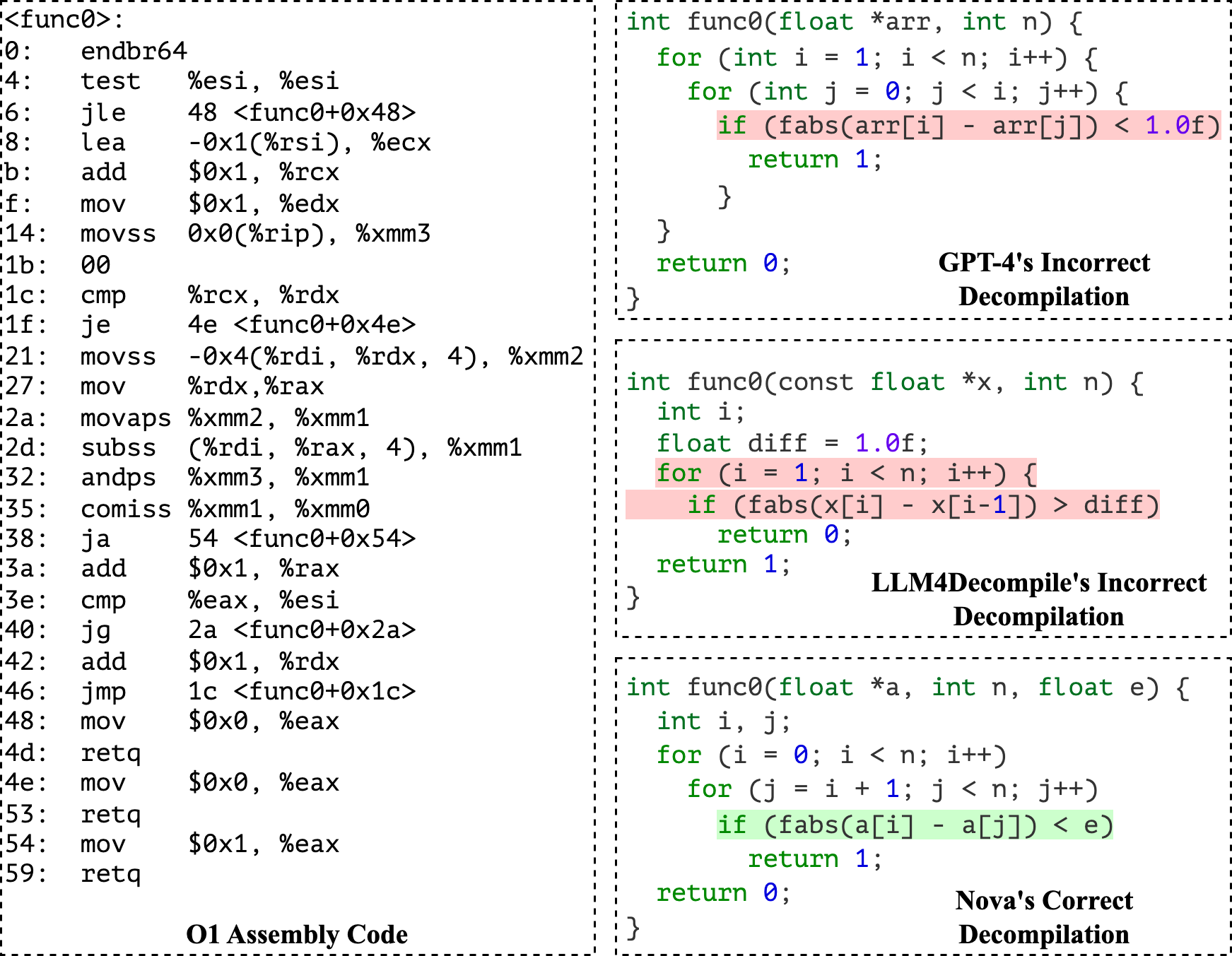}
    \caption{\ours{}-1.3B correctly decompiles HumanEval-Decompile task 0.}
    \label{fig:appendix_bcd_example_1}
\end{figure}

\begin{figure}[htp]
    \centering
    \includegraphics[width=0.75\linewidth]{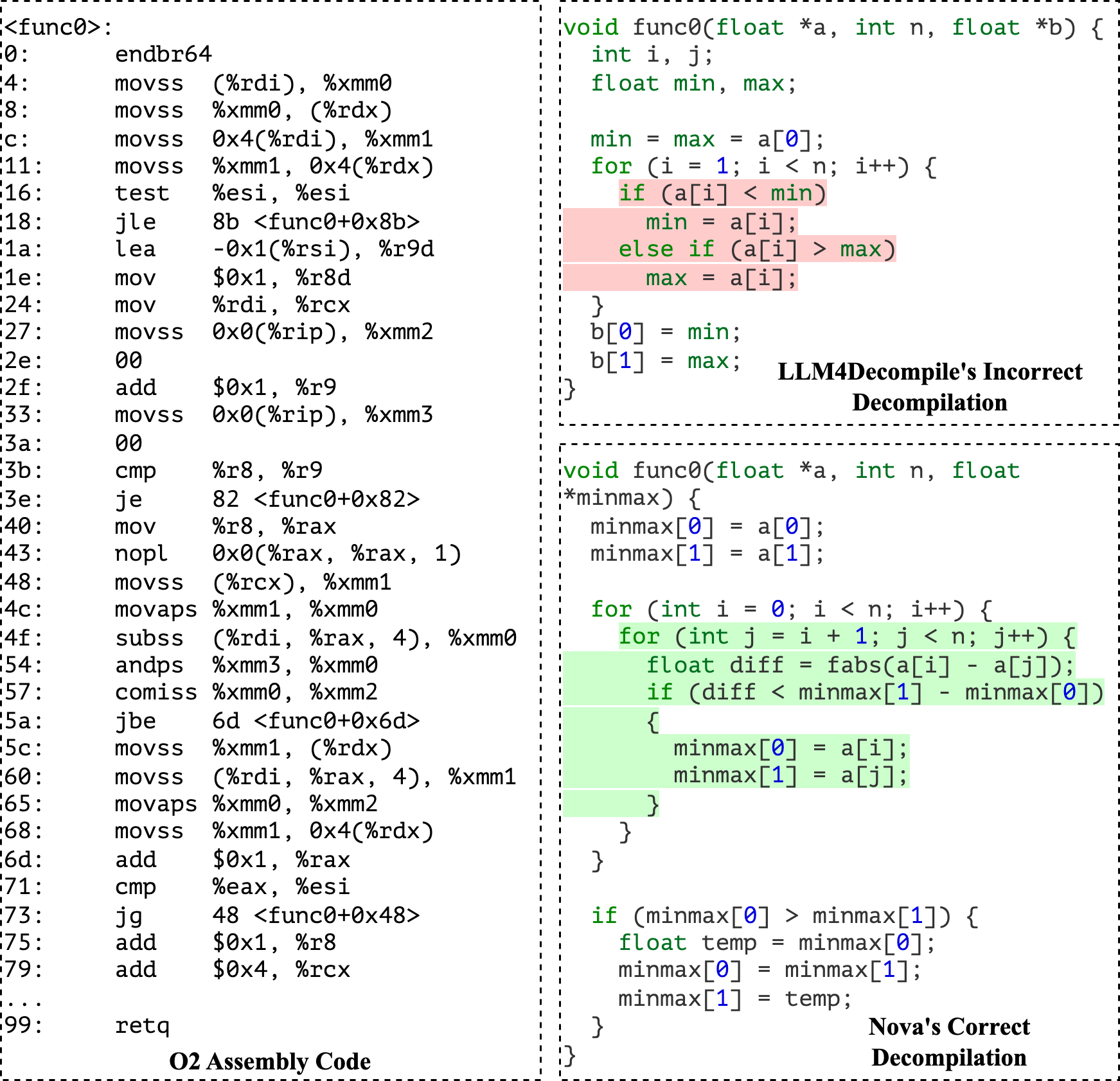}
    \caption{\ours{}-1.3B correctly decompiles HumanEval-Decompile task 20.}
    \label{fig:appendix_bcd_example_2}
\end{figure}

\subsection{Binary Code Decompilation Case Studies}
\label{sec:appendix_bcd_case_study}
Figure~\ref{fig:appendix_bcd_example_1} shows an example from HumanEval-Decompile (\texttt{task\_id} 0). Given the \code{O1}-optimized assembly code, GPT-4 fails to figure out the number of function arguments correctly, missing one important argument ``\code{float e}'', and thus produces wrong functionality in the decompiled code. LLM4Decompile-1.3B makes similar mistakes and also misses the inner nested \code{for} loop. \ours{}-1.3B correctly decompiles the assembly into source code, where the ground truth is checking if any two elements in the given list \code{*a} (with size \code{n}) are close to each other than a given threshold \code{e}.

Figure~\ref{fig:appendix_bcd_example_2} shows another more complex example, HumanEval-Decompile \texttt{task\_id} 20. \ours{}-1.3B correctly decompiles the source code, successfully figuring that the function is trying to find the two elements that are closest to each other in the given array \code{*a}, with the result stored in \code{minmax}.

\begin{table*}[htp]
    \scriptsize
    \centering
    \begin{tabular}{cc}
        \begin{minipage}{0.49\textwidth}
            \centering
            \caption{Ablation study with $K=50$.}
            \begin{tabular}{l|ccc}
                \toprule
                    Benchmarks & \nocl{} & \noha{} & \ours{}-1.3B \\
                \midrule
                    Binutils & 0.86 & \underline{0.88} & 0.87 \\
                    Curl & 0.84 & 0.87 & \underline{0.89} \\
                    ImageMagick & 0.79 & 0.80 & \underline{0.86} \\
                    SQLite & 0.80 & \underline{0.83} & 0.77 \\
                    OpenSSL & 0.90 & \underline{0.92} & 0.90 \\
                    Putty & 0.68 & 0.66 & \underline{0.72} \\
                \midrule
                    Avg. & 0.81 & 0.83 & \underline{0.84} \\
                \bottomrule
            \end{tabular}
            \label{tab:appendix_bcsd_ablation_k_50}
        \end{minipage} &
        \begin{minipage}{0.49\textwidth}
            \centering
            \caption{Ablation study with $K=100$.}
            \begin{tabular}{l|ccc}
                \toprule
                    Benchmarks & \nocl{} & \noha{} & \ours{}-1.3B \\
                \midrule
                    Binutils & 0.80 & \underline{0.82} & 0.79 \\
                    Curl & 0.84 & 0.84 & \underline{0.86} \\
                    ImageMagick & 0.70 & 0.72 & \underline{0.79} \\
                    SQLite & 0.74 & \underline{0.78} & 0.73 \\
                    OpenSSL & \uwave{0.89} & \uwave{0.89} & 0.88 \\
                    Putty & 0.59 & 0.60 & \underline{0.65} \\
                \midrule
                    Avg. & 0.76 & \uwave{0.78} & \uwave{0.78} \\
                \bottomrule
            \end{tabular}
            \label{tab:appendix_bcsd_ablation_k_100}
        \end{minipage} \\
        \multicolumn{2}{c}{} \\[0.2em]
        \begin{minipage}{0.49\textwidth}
            \centering
            \caption{Ablation study with $K=200$.}
            \begin{tabular}{l|ccc}
                \toprule
                    Benchmarks & \nocl{} & \noha{} & \ours{}-1.3B \\
                \midrule
                    Binutils & 0.71 & \underline{0.74} & 0.73 \\
                    Curl & 0.80 & 0.73 & \underline{0.83} \\
                    ImageMagick & 0.61 & 0.63 & \underline{0.73} \\
                    SQLite & 0.68 & \underline{0.71} & 0.68 \\
                    OpenSSL & 0.85 & \underline{0.87} & 0.84 \\
                    Putty & 0.53 & 0.53 & \underline{0.55} \\
                \midrule
                    Avg. & 0.70 & 0.70 & \underline{0.73} \\
                \bottomrule
            \end{tabular}
            \label{tab:appendix_bcsd_ablation_k_200}
        \end{minipage} &
        \begin{minipage}{0.49\textwidth}
            \centering
            \caption{Ablation study with $K=500$.}
            \begin{tabular}{l|ccc}
                \toprule
                    Benchmarks & \nocl{} & \noha{} & \ours{}-1.3B \\
                \midrule
                    Binutils & 0.62 & \uwave{0.65} & \uwave{0.65} \\
                    Curl & 0.67 & 0.71 & \underline{0.73} \\
                    ImageMagick & 0.46 & 0.51 & \underline{0.61} \\
                    SQLite & 0.61 & \underline{0.62} & 0.59 \\
                    OpenSSL & 0.77 & \underline{0.79} & 0.78 \\
                    Putty & 0.46 & 0.46 & \underline{0.47} \\
                \midrule
                    Avg. & 0.60 & 0.62 & \underline{0.64} \\
                \bottomrule
            \end{tabular}
            \label{tab:appendix_bcsd_ablation_k_500}
        \end{minipage} \\
    \end{tabular}
\end{table*}

\subsection{Binary Code Similarity Detection Ablation Study}
\label{sec:appendix_bcsd_ablation}
Table~\ref{tab:appendix_bcsd_ablation_k_50},~\ref{tab:appendix_bcsd_ablation_k_100},~\ref{tab:appendix_bcsd_ablation_k_200},~\ref{tab:appendix_bcsd_ablation_k_500} show the detailed ablation study results of BCSD. \ours{} wins on the most benchmarks when $K=100$ or $500$, and ties with \noha{} when $K=50$, or $200$.

\begin{figure*}[htp]
    \centering
    \includegraphics[width=0.8\linewidth]{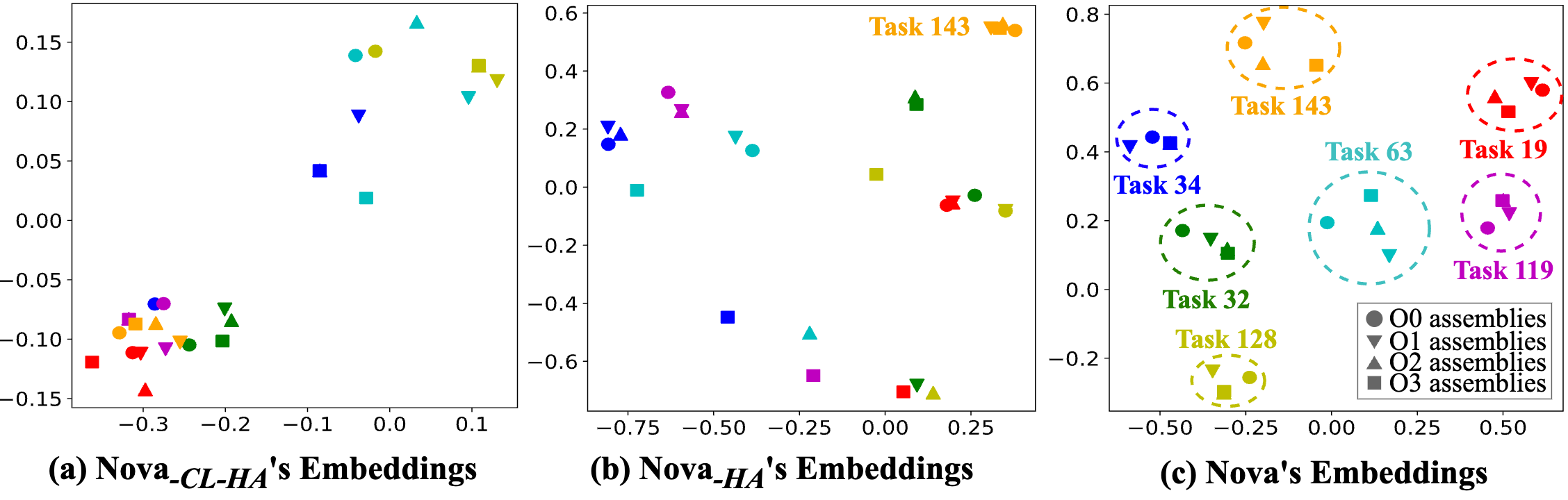}
    \caption{PCA of embeddings calculated by \nocl{}, \noha{}, and \ours{}.}
    \label{fig:embedding_pca}
\end{figure*}

\subsection{Additional Analysis of Embedding}
\label{sec:appendix_embedding_analysis}
Figure~\ref{fig:embedding_pca} shows the results of PCA of embeddings provided by \nocl{}, \noha{}, and \ours{}, on randomly sampled seven examples, where \ours{}'s embeddings are consistently more distinguishable by functionalities.

\begin{figure*}[htp]
    \centering
    \includegraphics[width=0.9\linewidth]{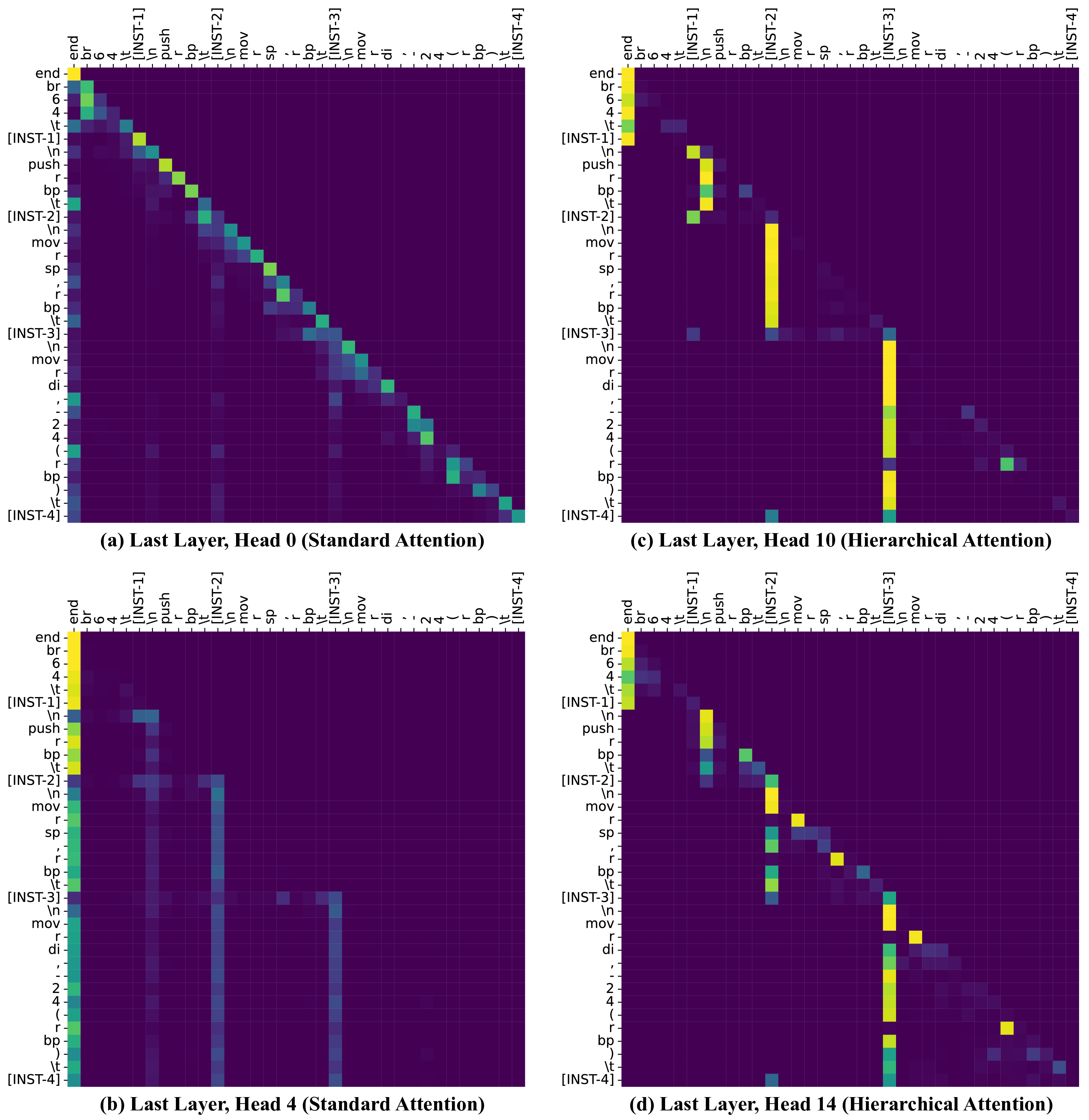}
    \caption{Comparison of attention distribution among standard and hierarchical heads.}
    \label{fig:attention-heads}
\end{figure*}

\subsection{Additional Analysis of Attention}
\label{sec:appendix_attention_analysis}

\begin{figure}[htp]
    \centering
    \includegraphics[width=0.5\linewidth]{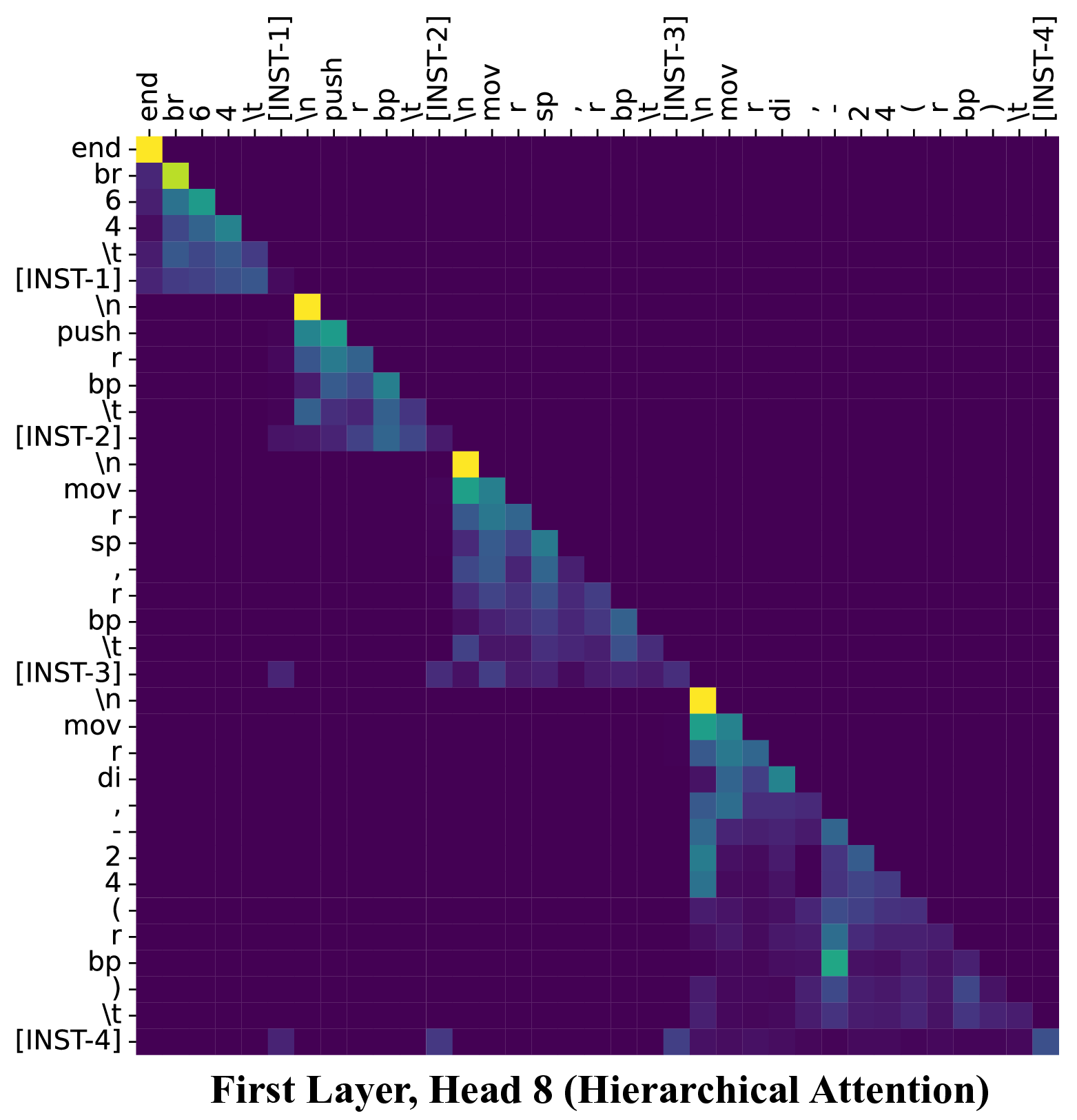}
    \caption{Learned per-instruction soft attention observed in the lower layers}
    \label{fig:attention-soft}
\end{figure}

Figure~\ref{fig:attention-heads} shows the visualizations of attention weights in the final transformer layer of two select heads with standard attention and two heads with learned hierarchical attention. Standard attention exhibits two typical patterns, namely diagonal attention (i.e. tokens attending to themselves or nearby tokens, shown in Figure~\ref{fig:attention-heads} (a)), and broad attention (i.e. a single token attending broadly to the entire sequence, shown in Figure~\ref{fig:attention-heads} (b)). In contrast, in \ours{}'s hierarchical attention, attention weights are allocated among distinct segments, each corresponding to an instruction (shown in Figure~\ref{fig:attention-heads} (c)), that focus on tokens comprising that instruction (e.g. opcodes and operands, shown in Figure~\ref{fig:attention-heads} (d), attentions are paid to ``\code{push}'', ``\code{mov}'', etc.).

Quantitatively, we have determined broad attention accounts for as much as 30\% of all attention in standard heads, especially in layers 1-8 (consistent with the findings of \citep{explain-bert}), whereas in \ours{}'s hierarchical attention, no more than 5\% or all attention is allocated to each instruction segment. This validates our goal of learning instruction-aware hierarchical attention in \ours{}.

In addition, in lower layers, we have observed attention weights to be softly distributed among tokens comprising each instruction (Figure~\ref{fig:attention-soft}), which suggests \ours{} initially models cross-relations among operation codes and operands in the first few layers, and later pools their summary representation into the \code{[INST]} token in the later layers.

\subsection{Limitations}
\rev{One limitation is that \ours{} is X86-specific, as we only collect X86 assembly corpus for pre-training. This design choice is mainly affected by two facts: (1) X86 assembly is used and explored in a wide range of binary tasks~\citep{jtrans, codeart, xu2023diemph, dirty} compared to other assembly languages, and (2) computation limitations. However, the proposed techniques are independent of X86 assembly. Low information density and compiler optimization are the common challenges of most assembly languages such as X86, ARM, and MIPS. The proposed techniques can be applied to the future development of multi-lingual assembly LLMs. Another potential limitation is the scale of models. We develop \ours{}-1.3B and \ours{}-6.7B and demonstrate their advancement in assembly code decompilation and encoding. Developing larger \ours{} models is promising, but remains as future work due to limited computing resources.} 

\end{document}